\documentclass[twocolumn,prl,aps,floatfix,superscriptaddress]{revtex4-2}
\usepackage{amsmath}
\usepackage{amssymb}
\usepackage{graphicx}

\usepackage{color}
\usepackage{xcolor}
\usepackage{amssymb}
\usepackage{amsmath}
\usepackage{bm}
\usepackage{hyperref}
\usepackage{graphicx}
\usepackage{amsmath,amssymb,bm,epsfig,color}
\usepackage{mathtools}

\newcommand{\vf}{\varphi}

\makeatletter
\usepackage{bm}\usepackage{epsfig}\usepackage{color}\usepackage{siunitx}
\usepackage{subfigure}

\makeatother

\begin{document}
\title{Interacting type-II semi-Dirac quasiparticles}
\author{Mohamed M. Elsayed}
\affiliation{Department of Physics, University of Vermont, Burlington, Vermont 05405, USA}
\author{Taras I. Lakoba}
\affiliation{Department of Mathematics and Statistics, University of Vermont, Burlington, Vermont 05405, USA}
\author{Valeri N. Kotov}
\affiliation{Department of Physics, University of Vermont, Burlington, Vermont 05405, USA}

\begin{abstract}
 Type-II semi-Dirac fermions in two dimensions have been proposed to describe topologically nontrivial low-energy excitations in titanium/vanadium oxide heterostructures. These quasiparticles appear at the merger of three Dirac cones, resulting in a 
 non-zero  
 Berry phase. We find, by employing Hartree-Fock, renormalization group, and Random Phase Approximation (RPA) techniques, that the spectrum is very sensitive to long-range electron-electron interactions and can undergo a profound transformation. Our results indicate that at the topological phase boundary, long-range correlations stabilize a hybrid electronic phase displaying both Dirac and type-II semi-Dirac qualities, with physical characteristics exhibiting continuously varying critical exponents as a function of the Fermi energy; for example Landau levels in a magnetic field vary with the energy scale: $|\varepsilon_n(B)|\sim (nB)^{1/2} \rightarrow  (nB)^{3/4},  n\in \mathbb{N}_0$. The quasiparticle spectrum evolves, driven by interactions, from anisotropic Dirac dispersion at the lowest energies, towards the characteristic type-II semi-Dirac boomerang shape as the energy increases.  The corresponding density of states concomitantly varies between linear and power one third ($\rho(\varepsilon) \sim |\varepsilon| \rightarrow |\varepsilon|^{1/3}$). The crossover scale is controlled by the interaction strength  $\alpha = e^2/(\hbar v)$ and the specifics of the effective interacting Hamiltonian.
 \end{abstract}
\maketitle

\begin{figure*}
            \centering
            \begin{minipage}{0.33\textwidth}
                 \includegraphics[width=1\textwidth]{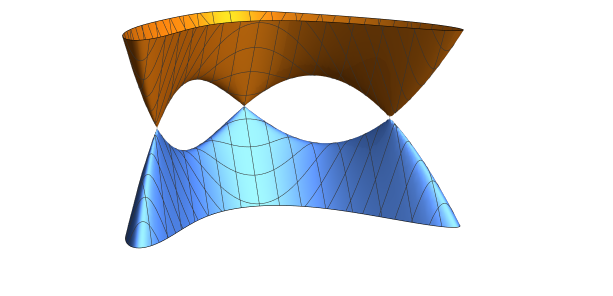}
                 \vspace{0.2cm}
                \hspace{-0.6cm} (a)
        \end{minipage}   
        \hspace{-0.2cm}
        \begin{minipage}{0.33\textwidth}
                 
                 \includegraphics[width=1\textwidth]{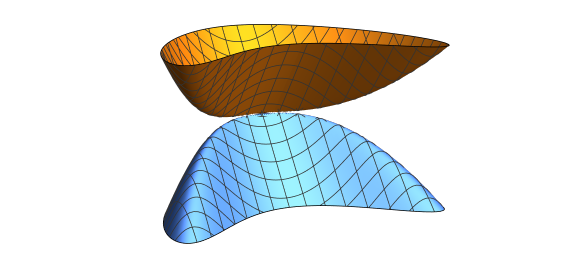}
                 \vspace{0.2cm}
                \hspace{-0.2cm} (b)
        \end{minipage}
        \begin{minipage}{0.33\textwidth}
                 \includegraphics[width=1\textwidth]{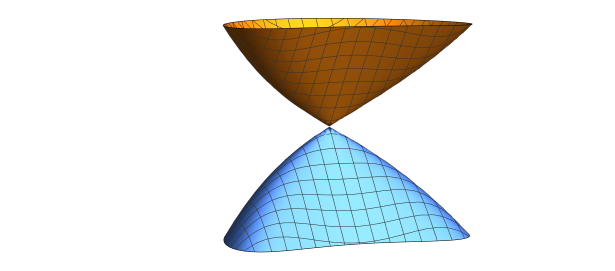}
                 \vspace{0.2cm}
                \hspace{0.4cm} (c)
        \end{minipage} 
\caption{Low-energy electronic dispersion corresponding to the Hamiltonian in Eq.(\ref{ham}) under a small perturbation $\Delta$, such that $\mathcal{H}=\left(g_1 k_{x}^{2} - v k_y +\Delta\right)\hat{\sigma}_{x}+g_2 k_{x}k_{y}\hat{\sigma}_{y}$. (a) $\Delta<0$, leading to the presence of three Dirac cones located at $(0,\Delta/v),(\pm\sqrt{-\Delta/g_1},0)$. (b) $\Delta=0$: The three cones coalesce at the origin, giving rise to a type-II semi-Dirac point with a finite Berry phase. (c) $\Delta>0$: The band crossing is shifted to $k_y=\Delta/v$. At the smallest momenta the spectrum is linear, but exhibits type-II semi-Dirac behavior at larger energies.}
\label{fig:intro} 
\end{figure*}

{\it Introduction---}
Semi-Dirac fermions are a hybrid low-energy excitation, exhibiting the qualities of both massless Dirac particles and massive Galilean particles at once. More precisely, they disperse linearly in a given direction in momentum space, and quadratically in the orthogonal direction. This is most easily understood as resulting from the merger of two Dirac cones at the critical point of a semimetal-insulator transition, whereby the Berry phases annihilate and the critical semi-Dirac spectrum emerges \cite{montambauxmerging,montambaux2009universal}. The fundamental anistropy gives rise to unconventional phenomena in a variety of contexts, such as: charge \cite{carbotte2019,oriekhov2022} and thermoelectric \cite{mawrie2019,vargiamidis2025} transport; behavior in strong magnetic fields \cite{sinha2022,asafov2024}; optical effects and surface interaction with light \cite{Zhou2021,cheng2025}; edge states and topological behavior in modified models with broken symmetries \cite{mondal2022,marta2025}; and anisotropic collective excitations such as plasmons, excitons, and spinons \cite{arafat2024,rossharvey2025,chatterjee2026,giri2025}. Furthermore, there are other possible flavors of semi-Dirac fermions. Higher order generalizations have attracted theoretical study, where the quadratic dispersion is generalized to an arbitrary even power law \cite{elsayed2025,roy2018}. Recently there has been a proposed lattice model realization for the most basic non-trivial case in this class of Hamiltonians, namely the quartic semi-Dirac fermions \cite{ElsayedQuartic}. 

Moreover, there is an even more exotic, and less explored, species of semi-Dirac fermions that are not topologically trivial. These excitations emerge at a critical point where not two but three Dirac cones coalesce, producing a finite Berry phase and what has been dubbed a type-II semi-Dirac spectrum \cite{huang2015}. The dispersion retains its semi-Dirac character in the sense that it is still linear and quadratic along the given axes, but exhibits a different admixture of momentum components for an arbitrary direction of propagation \cite{huang2015,Pardo2010}. This symmetry reduced model was first proposed to reconcile the finite Chern number and semi-Dirac behavior of quasiparticles in titanium/vanadium oxide heterostructures \cite{huang2015}. However, it may also be realized in more conventional platforms such as bilayer graphene, by leveraging interlayer sliding \cite{liao2025berry,Son2011}. Such systems have been shown to exhibit a rich variety of topological phases that are accessible via fine tuning with light-matter interactions \cite{Chen2022}. In addition, the peculiar anisotropy leads to highly unusual optical and transport properties \cite{Xiong2023,liao2025berry}. Topological semi-metallic phases have also been analyzed in detail both experimentally and theoretically in ZrSiS, reporting a measurement of the signature Landau level scaling with magnetic field \cite{shao2024}.

Electron-electron interactions are known to have pronounced effects in ordinary semi-Dirac fermions \cite{Dietl2008,Goerbig2017,deGail2012,kotov2021,montambauxmerging}. Short-range interactions have been shown to stabilize modulated ordered phases in the charge, spin, and superconducting sectors around a quantum critical point \cite{Uchoa2017,uryszek2019}. On the other hand, the long range Coulomb interaction produces two low-energy regimes in which there is unusually strong spectrum renormalization \cite{elsayed2025,isobe2016}. Motivated by these results, we study the spectrum renormalization of the topological type-II semi-Dirac fermions due to the long range, and RPA-screened, Coulomb interaction. We have found that,  remarkably, the interacting system hosts massless Dirac quasiparticles at the lowest energies, smoothly transitioning to the free fermionic behavior with increasing energy. This is captured in the exponent $\eta$ of the density of states ($\rho(\varepsilon)\sim\varepsilon^{\eta(\varepsilon)}$) as a function of energy varying from 1 to 1/3 on a sigmoid. The Fermi surface geometry evolves from convex to concave in accordance with the changing $\eta$. 
As a consequence, many physical properties and observables will have a continuously evolving dependence on the energy scale of interest. Given that Coulomb interactions are invariably present in systems of charged particles, our results can explain the appearance of the low-energy linear behavior in Ref.\cite{huang2015}. Moreover, our study implies that a measurement or calculation of the coefficient of the linear term in the Hamiltonian along the massive direction can be used as a diagnostic tool to estimate the effective interaction strength in such systems. As an illustration, we model the quasiparticles in the candidate material (TiO$_2$)$_5$(VO$_2$)$_3$ as interacting type-II fermions at the topological phase boundary, using parameter values from Ref.\cite{huang2015}. By comparing coefficients of the low-energy linear term along the massive direction, we make a rough estimate of the effective interaction strength in the material, which indeed lies in the weak coupling regime. Intriguingly, a similar change in Fermi surface shape (driven by temperature) in the 2D semimetal WTe$_2$ has been observed to lead to a spectacular reversal in the sign of the nonlinear current response under crossed electric and magnetic fields \cite{he2019}. Even though our work is concerned with zero temperature, interaction-driven effects, the qualitatively identical crossover prompts us to  expect similarly novel magnetotransport, and possibly other unusual phenomena.

\begin{figure}
\centering
 \includegraphics[width=1\linewidth]{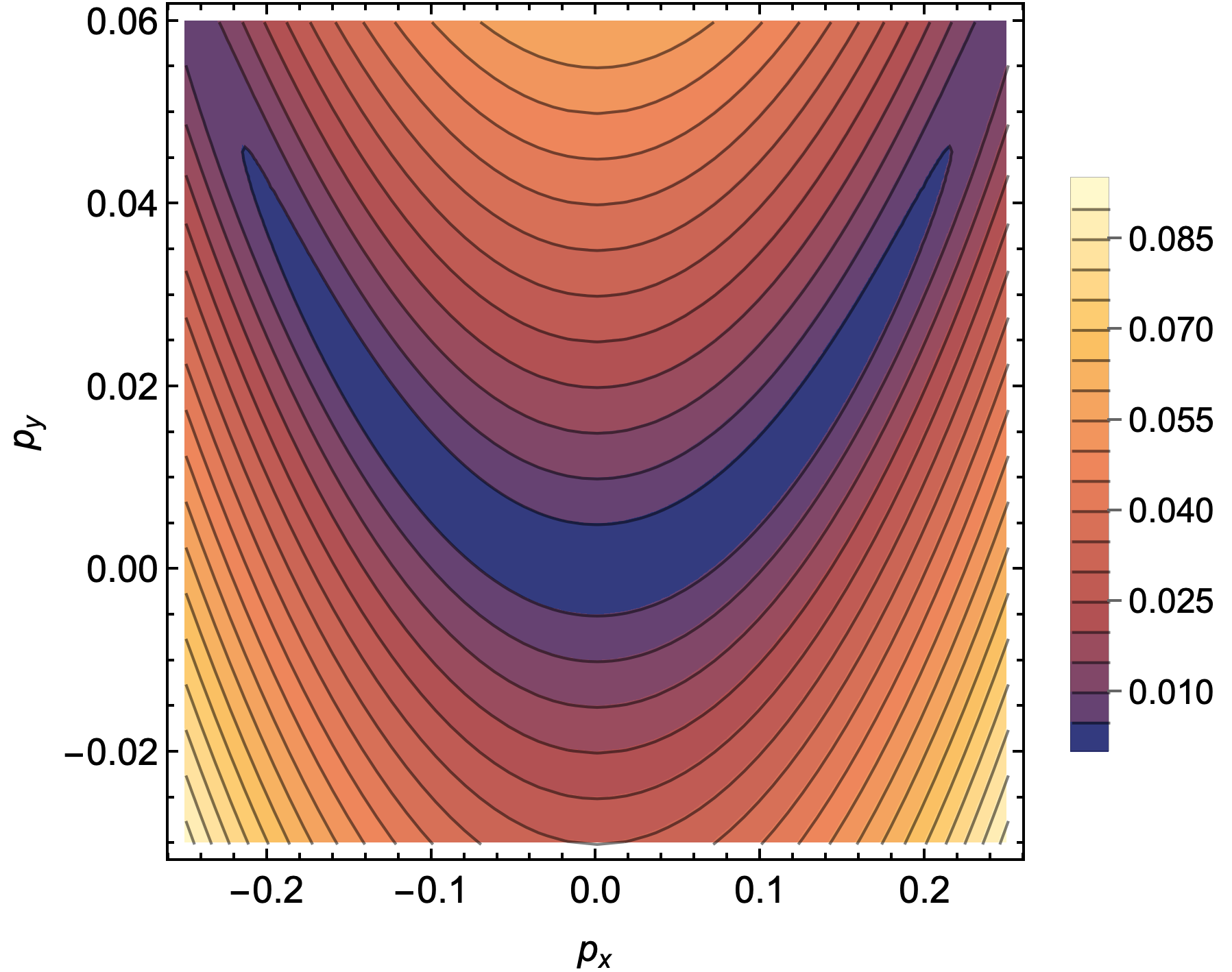}
    \caption{Contour plot of the critical low-energy spectrum Eq.(\ref{spec_rescaled}), illustrated schematically in Fig. \ref{fig:intro}b). Energies are measured in units of $\varepsilon_0$ and momenta in units of $v/g_1$. We take $g=1/2$, similar to the cross anisotropy in the candidate material. The isoenergy curves have a characteristic boomerang shape.}
  \label{fig:isosurf}
\end{figure}

We consider the Hamiltonian describing type-II semi-Dirac fermions \cite{huang2015}:
\begin{equation}
\mathcal{H}(\mathbf{k})=\left(g_1 k_{x}^{2} - v k_y\right)\hat{\sigma}_{x}+g_2 k_{x}k_{y}\hat{\sigma}_{y},
\label{ham}
\end{equation}
where the coupling $g_1$ is the inverse effective mass, $v$ is the velocity in the massless direction, and $g_2$ is the cross anisotropy that characterizes the type-II nature of the particles. Such excitations can arise at the critical point of the topological transition shown in Fig. \ref{fig:intro}. 
We rescale the momenta, defining
$p_{x,y}=(g_1/v)k_{x,y}$, 
and introduce the energy scale $\varepsilon_0=v^2/g_1$ to arrive at the electronic dispersion 
\begin{equation}
\varepsilon(\mathbf{p})=\pm \varepsilon_0\sqrt{\left(p_{x}^{2} - p_y\right)^2+(gp_{x}p_{y})^2}, 
\label{spec_rescaled}
\end{equation}
where we define $g\equiv g_2/g_1$. A plot of the spectrum is shown in Fig.\ref{fig:isosurf} using a representative $g=1/2$, similarly to the value of $g=0.41$ in (TiO$_2$)$_5$(VO$_2$)$_3$ \cite{huang2015}.

{\it Energy variables and density of states---} When calculating diagrams and other quantities of interest in systems with such anisotropy, it is useful to recast into energy-angle variables which we define as follows:
\begin{eqnarray}
&p_x^2 - p_y = (\varepsilon/\varepsilon_0) \cos \varphi,&\nonumber \\
&gp_xp_y=(\varepsilon/\varepsilon_0)\sin \varphi.&
\end{eqnarray}
The solution of these equations in the low-energy limit ($\varepsilon/\varepsilon_0 \ll 1$) is
\begin{eqnarray}
&&p_x = \text{sgn}(\sin{\varphi}) \left(\frac{E|\sin\varphi|}{g}\right)^{1/3} + \mathcal{O}(E^{2/3}), \nonumber \\
&&p_y = \left(\frac{E|\sin\varphi|}{g}\right)^{2/3}-E\cos{\varphi} + \mathcal{O}(E^{4/3}), \nonumber \\
&&{\mbox{where}} \ \ \  0\leq \varphi \leq 2\pi,  \ \ E \equiv \frac{\varepsilon}{\varepsilon_0}.
\label{cardano}
\end{eqnarray}
In this limit we calculate the Jacobian of the transformation, immediately yielding the  
density of states as a function of energy $\varepsilon$
\begin{equation}
\rho(\varepsilon)  = \iint \frac{d^2k}{(2\pi)^2} \delta \left(\varepsilon -\varepsilon(\mathbf{k})\right)   
=\frac{1}{6\pi^{3/2}}\frac{\Gamma(1/6)}{\Gamma(2/3)}\left(\frac{E}{g_1^{\,2}\, g_2}\right)^{1/3},
\end{equation}
with the numerical coefficient amounting to $0.123$.
We find the non-linear transformation Eq.~\eqref{cardano} to be particularly effective in rendering the problem analytically tractable in the low-energy limit.

{\it Perturbation theory and Renormalization Group---}We take into account the non-retarded Coulomb interaction
$V({\bf k})=2\pi e^{2}/|{\bf k}|$ to first order in perturbation
theory. The effective dimensionless coupling that controls the perturbative expansion is $\alpha=e^2/v$ ($\hbar=1$). The self-energy
 is
 $\hat{\Sigma}(\mathbf{k})=i/(2\pi)^3\int_{-\infty}^{\infty}\text{d}\nu\int\text{d}^{2}k'\,\hat{G}(\mathbf{k'},\nu)V(\mathbf{k}-\mathbf{k'})$,
where
$\hat{G}^{-1}(\mathbf{k},\nu)=\nu-\mathcal{H}(\mathbf{k}) + i0^{+}{\mbox{sign}}(\nu)\label{G}$
defines the fermionic Green's function. The frequency integral can be easily
evaluated: 
\begin{equation}
\hat{\Sigma}({\bf k})=\frac{1}{2}\int\frac{\text{d}^{2}k'}{(2\pi)^{2}}\frac{2\pi e^{2}}{|{\bf k}-{\bf k'}|}\frac{1}{|\varepsilon({\bf k'})|} \mathcal{H}(\mathbf{k'}).\label{se}
\end{equation}
We calculate corrections: $\Sigma_v$ to the quasiparticle velocity, $\Sigma_{g_1}$ to the inverse mass, and $\Sigma_{g_2}$ to the cross anisotropy, which result in the following universal Hamiltonian
\begin{eqnarray}
\tilde{\mathcal{H}}(\mathbf{k})&=&\left [g_1 (1+\Sigma_{g_1})k_{x}^{2} - v (1+\Sigma_{v})k_y\right ]\hat{\sigma}_{x} \nonumber \\
&&+ g_2 (1+\Sigma_{g_2}) k_{x}k_{y}\hat{\sigma}_{y},
\label{ham_pert}
\end{eqnarray}
omitting terms dependent on the high-energy cutoff $\Lambda$. For example the correction $\Sigma_v$ may be extracted by considering the asymptotic low-energy behavior of
\begin{eqnarray}
    \Sigma_v(\omega) &=& \frac{\alpha}{12\pi}\int_{\omega}^{\Lambda}\frac{\text{d}E}{E} I_{v}(E), \nonumber \\
    I_{v}(E) &=& 2\int_{0}^{\pi}\text{d}\varphi \frac{E^{1/3}}{[\sin{\varphi}]^{2/3}} \nonumber \\
 &\times& \frac{(E^{2/3}[\sin{\varphi}]^{2/3}-E \cos{\varphi})^2}{\left(E^{2/3}[\sin{\varphi}]^{2/3}+(E^{2/3}[\sin{\varphi}]^{2/3}-E \cos{\varphi})^2\right)^{3/2}} \nonumber 
\end{eqnarray}
which produces a leading logarithmic divergence $\Sigma_v=(\alpha/\pi )\ln(\Lambda/\omega)$. Details of the computation for the parameters $g_{1,2}, v$ are outlined in the Supplementary Material (SM). We  have the following one-loop perturbation theory results: 
\begin{eqnarray}
v(\omega)&=&vL(\omega),\,\, g_{1,2}(\omega)=g_{1,2}L(\omega),\nonumber\\
L(\omega) &=& \left(1+ \frac{\alpha}{\pi}\ln(\Lambda/\omega) \right)=(1+\Sigma_{v,g_1,g_2}),
\label{v-leading}
\end{eqnarray}
where the infrared $\omega$ will be taken to be on-shell 
$\omega\equiv\varepsilon(\bf{k})$ 
in subsequent calculations i.e. follows the form of the bare dispersion.

Next, we write down and integrate the corresponding Renormalization Group (RG) equations, 
  where we introduce ${\it \ell} = \ln(\Lambda/\omega)$:
$dv/d{\it \ell}=v(\ell) \alpha(\ell)/\pi=e^2/\pi, \  dg_1/d{\it \ell}=\alpha(\ell) g_1(\ell)/\pi,  \  dg_2/d{\it \ell}=\alpha(\ell) g_2(\ell)/\pi$. From the first equation it follows, since the electric charge does not renormalize, that the solution for $v(\ell)$ is trivial returning the perturbative result Eq.~\eqref{v-leading}, and $\alpha$ only renormalizes via its dependence on $v$: $\alpha(\omega)=\alpha/\left[1+\frac{\alpha}{\pi}\ln(\Lambda/\omega)\right]$ . This indicates that there are no higher-order logarithmic divergences for the RG to resum. It is is easy to see that the same applies for $g_{1,2}$, and the equations integrate to what are now our final results in Eq.~\eqref{v-leading}. Similarly to graphene \cite{Kotov2012} and ordinary semi-Dirac fermions, the coupling constant is marginally irrelevant. However, in contrast to type-I semi-Dirac fermions where the presence of a characteristic $\log^2$ divergence in $g_1$ at first order generates a higher-order resummation \cite{kotov2021,elsayed2025}, the type-II couplings remain unchanged under RG. 

{\it Effective Hamiltonian---}
In addition to the logarithmic corrections, there appear two crucial  cutoff-dependent terms in the self-energy. The zero-momentum contribution reads
\begin{eqnarray}
    \hat{\Sigma}(0)&=&\alpha\ \left( \frac{\varepsilon_0}{4\pi}\int \frac{\text{d}^2 p}{\sqrt{p_x^2+p_y^2}}\frac{p_x^2-p_y}{\sqrt{(p_x^2-p_y)^2+(g p_x p_y)^2}}\right)\hat{\sigma}_x \nonumber \\
    &\equiv& \alpha \Delta\, \hat{\sigma}_x,
\end{eqnarray}
and the corresponding $\hat{\sigma}_y$ component vanishes by parity. The generation of a constant, zero-momentum, term induced by interactions in the effective Hamiltonian, is a characteristic feature of semi-Dirac fermions of all types.  Moreover, expanding the Coulomb potential for small external momenta $V({\bf k}-{\bf k'})\approx\frac{2\pi e^{2}}{k}\left(1+\frac{{\bf k}\cdot{\bf k'}}{k^{2}}\right)$,
we find that the only (unaccounted for) term from Eq.~(\ref{se}) that does not vanish by parity is the one $\propto k_x \hat{\sigma}_y$, generating the following contribution
\begin{eqnarray}
&&\alpha \left( \frac{v}{4\pi}\int \frac{\text{d}^2 p}{(p_x^2+p_y^2)^{3/2}}\frac{g p_x^2 p_y}{\sqrt{(p_x^2-p_y)^2+(g p_x p_y)^2}}\right) k_x \hat{\sigma}_y \nonumber \\
&&\equiv \alpha ck_x\, \hat{\sigma}_y.
\end{eqnarray}
Assembling the corrections together we arrive at the interacting effective Hamiltonian:
\begin{eqnarray}
\mathcal{H}(\mathbf{k})&=&\left (g_1L(\omega) k_{x}^{2} - vL(\omega) k_y \! + \! \alpha \Delta \right )\hat{\sigma}_{x} 
\nonumber \\
&& +\left ( g_2  L(\omega) k_{x}k_{y} + \alpha c k_{x}  \right )\hat{\sigma}_{y}.
\label{ham_eff}
\end{eqnarray}
\begin{figure}
\centering
 \includegraphics[width=1\linewidth]{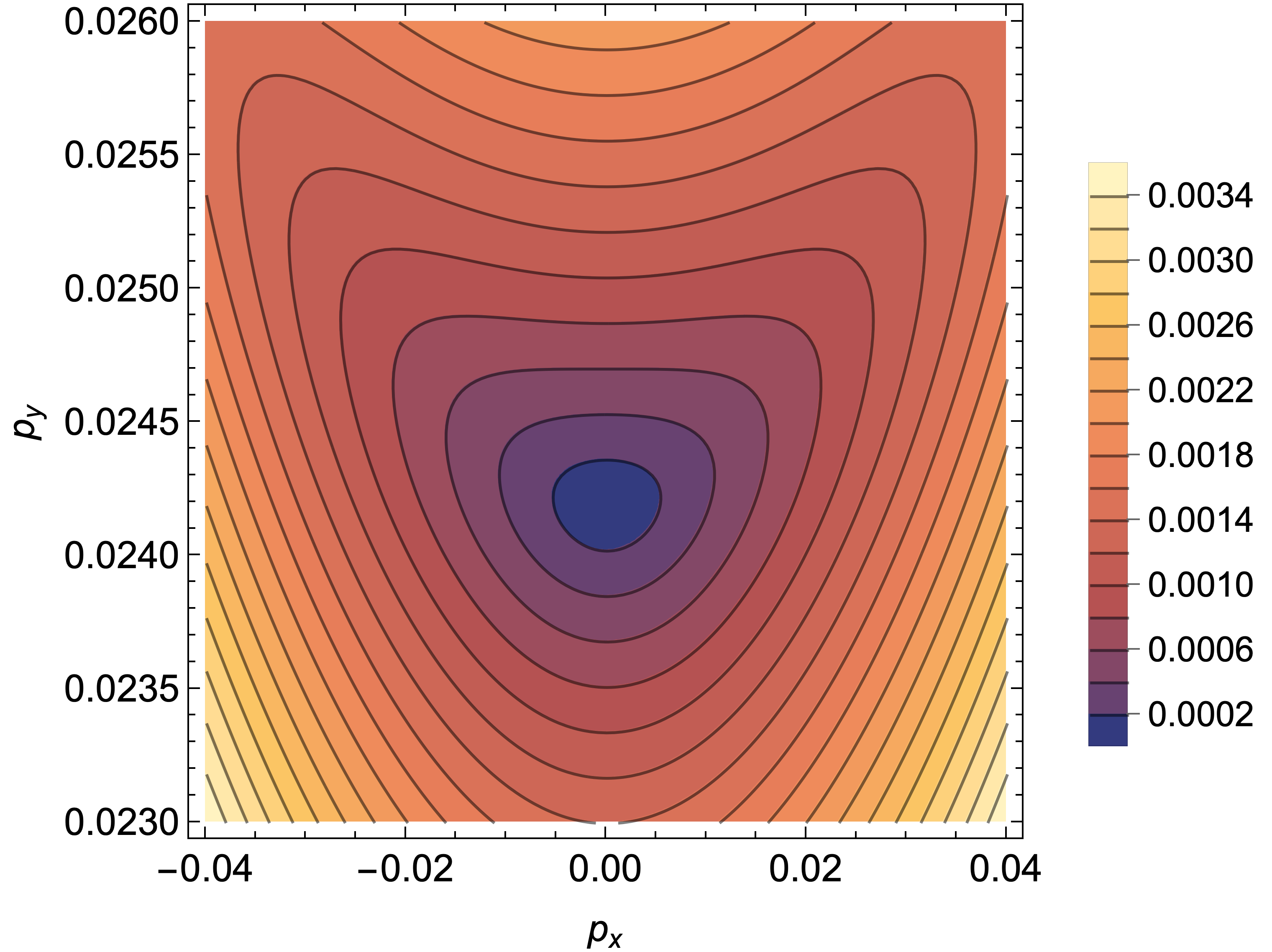}
   \caption{Type-II semi-Dirac spectrum renormalized by the bare Coulomb interaction at the one-loop Hartree-Fock level for $\alpha=0.2$ and $g=1/2$. The convex lowest energy contours indicate an anisotropic Dirac cone, with concave semi-Dirac behavior appearing at higher energies corresponding to Eq.~\eqref{crossover}. Noting the relatively narrow range of $p_y$ considered in the plot, it is clear that $p_x$ controls the crossover energy scale at which the Fermi surface changes geometry and the qualitative nature of quasiparticles transforms. An ultraviolet momentum cutoff  $k_{\Lambda}=1.1$ on a disk region, in units of $v/g_1$, is used to produce a numerical fit of Eq.~(\ref{se}) that returns the values $\Delta/\varepsilon_0=1.88/4\pi$ and $c/v=1.39/4\pi$.}
  \label{fig:isosurfHF}
\end{figure}
\begin{figure}
\centering
 \includegraphics[width=0.9\linewidth]{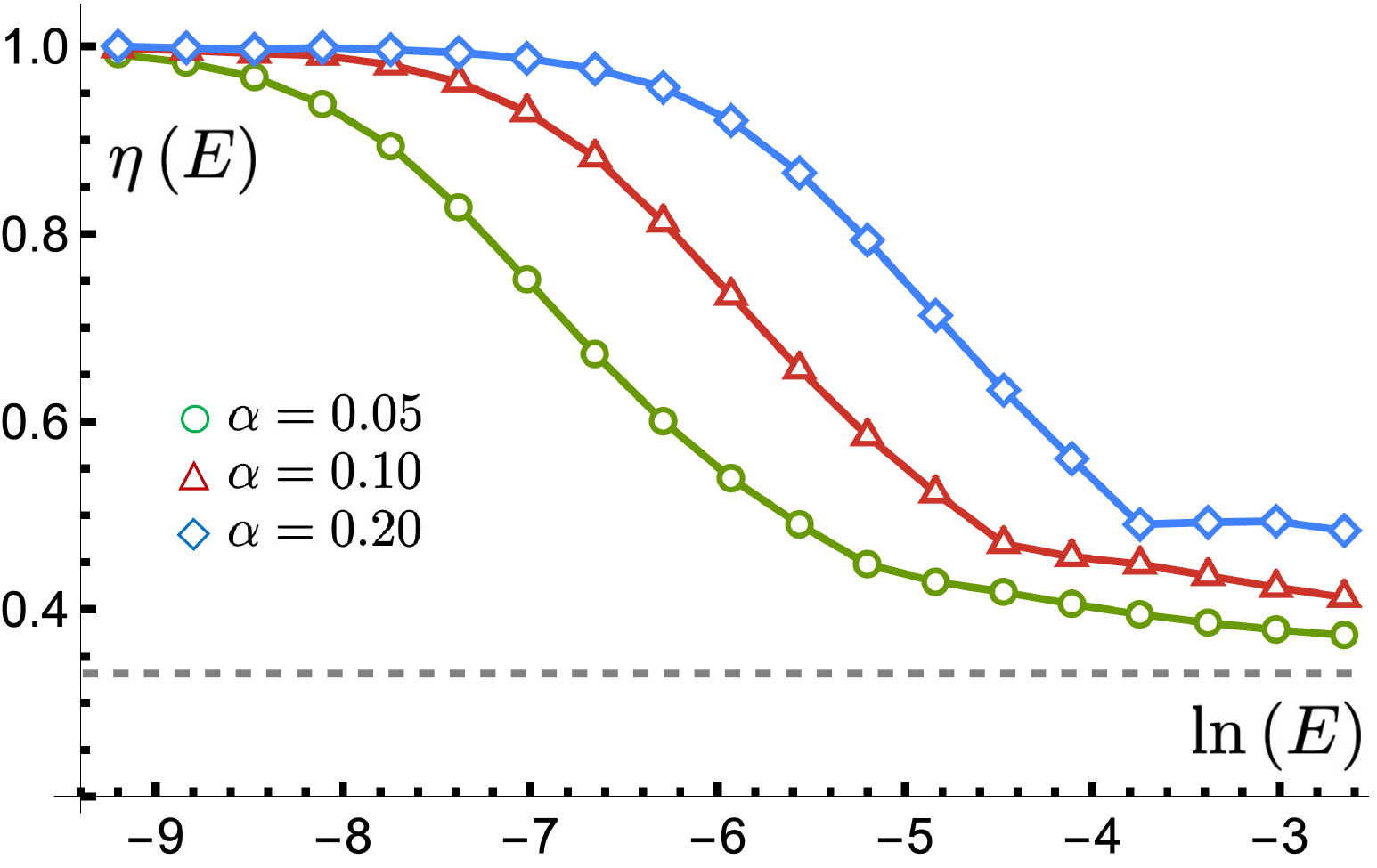}
   \caption{Evolution of the power $\eta(E)$ in the density of states $\rho(E)\sim E^{\eta(E)}$ for $g=1/2$ under different interaction strengths $\alpha=0.05,0.10,0.20$ in the Hartree-Fock approximation. The smooth transition from linear Dirac behavior $\eta=1$ at the smallest energies, approaching type-II semi-Dirac behavior $\eta=1/3$ (dashed line) with increasing energy, captures the variation in the interacting spectrum shown in Fig. \ref{fig:isosurfHF}.}
  \label{fig:dosHF}
\end{figure}
Henceforth $v,g_1,g_2,\alpha$ will denote the bare couplings. We find $\Delta >0$ which implies a shift of the dispersion in the $k_y$ direction, meaning that the band crossing is now at $k_y=\alpha\Delta/v$ (whereas $\Delta <0$ would have meant the presence of three Dirac cones, see Fig. \ref{fig:intro}). Measuring momenta from the new semi-Dirac point, the Hamiltonian assumes the same form as Eq.~(\ref{ham}), but with the logarithmically enhanced parameters in Eq.~(\ref{v-leading}), and an additional interaction-driven linear correction 
\begin{equation}
   \alpha\left(\frac{g_2}{v}\Delta+c\right)k_x \, \hat{\sigma}_y,
   \label{effectivelinearterm}
\end{equation}
which indicates that the dispersion is effectively anisotropic Dirac at small momenta. As shown in Fig. \ref{fig:isosurfHF} the spectrum smoothly transitions from linear Dirac to type-II semi-Dirac behavior with increasing energy. To capture this crossover, we model the density of states for the interacting fermions as  $\rho(\varepsilon) \sim \varepsilon^{\eta(\varepsilon)}$. In order to compute $\eta$, we numerically evaluate the area enclosed by cyclotron orbits in momentum space as a function of energy. A discretized differentiation returns the density of states, from which the exponent can be easily extracted. Results for varying interaction strength are shown in Fig. \ref{fig:dosHF}. The crossover from linear to semi-Dirac excitations is clearly reflected in the decrease of $\eta$ from 1 to the saturation at 1/3. The onset is controlled by an interaction dependent energy scale corresponding to 
 \begin{equation}
\vert k_x^* \vert\approx \frac{\alpha[(g_2/v)\Delta+c]}{g_1\,L\left(\omega(k_x^*)\right)}
\label{crossover},
\end{equation}
the regime where the quadratic term $\sim k_x^2$ in Eq.~(\ref{ham_eff}) begins to dominate over the linear $k_x$ term in Eq.~(\ref{effectivelinearterm}). At small energies the renormalization (increase) of $g_1$ favors the semi-Dirac spectrum, and suppresses the linear regime. In contrast, the running of $g_1$ under Coulomb interactions in ordinary semi-Dirac fermions gives rise to a linear spectrum for a large window of intermediate energies, with the renormalized semi-Dirac behavior present at the lowest energies \cite{elsayed2025}.

To showcase the predictive power of our results, we will attempt a rough estimate of the effective interaction strength in (TiO$_2$)$_5$(VO$_2$)$_3$, for which the band structure is well-known to be type-II semi-Dirac with a small low-energy linear regime. Modeling the low-energy excitations of the material as interacting fermions at the critical point, the aim is to deduce the effective interaction strength that would reproduce the coefficient of the linear term found in the physical system. We assume  parameters $g_1=0.256$, $g_2=0.1044$, and $\vert v \vert =0.3929$, identical to the material \cite{huang2015}; and numerically evaluate $\Delta =\left[\varepsilon_0/(4\pi)\right]2.244$, and $c=\left[v/(4\pi)\right]0.892$ on a square region with a high-energy cutoff $\vert p_{\Lambda}^{x,y} \vert =1.1$. Comparing Eq.~(\ref{effectivelinearterm}) to the term present in the (TiO$_2$)$_5$(VO$_2$)$_3$ Hamiltonian: $-0.02044 k_x\,\hat{\sigma}_y$, we arrive at an effective interaction strength of $\alpha_{\text{eff}}=0.36$, which is well within the weak coupling regime. In light of the conclusion in Ref.\cite{huang2015} that the material is an "anisotropic linear Dirac system that is so close to a
type-II semi-Dirac behavior that it can hardly be distinguished
from it", we feel that our characterization of the quasiparticles as weakly interacting fermions at criticality is justified.

\begin{figure}
\centering
 \includegraphics[width=1\linewidth]{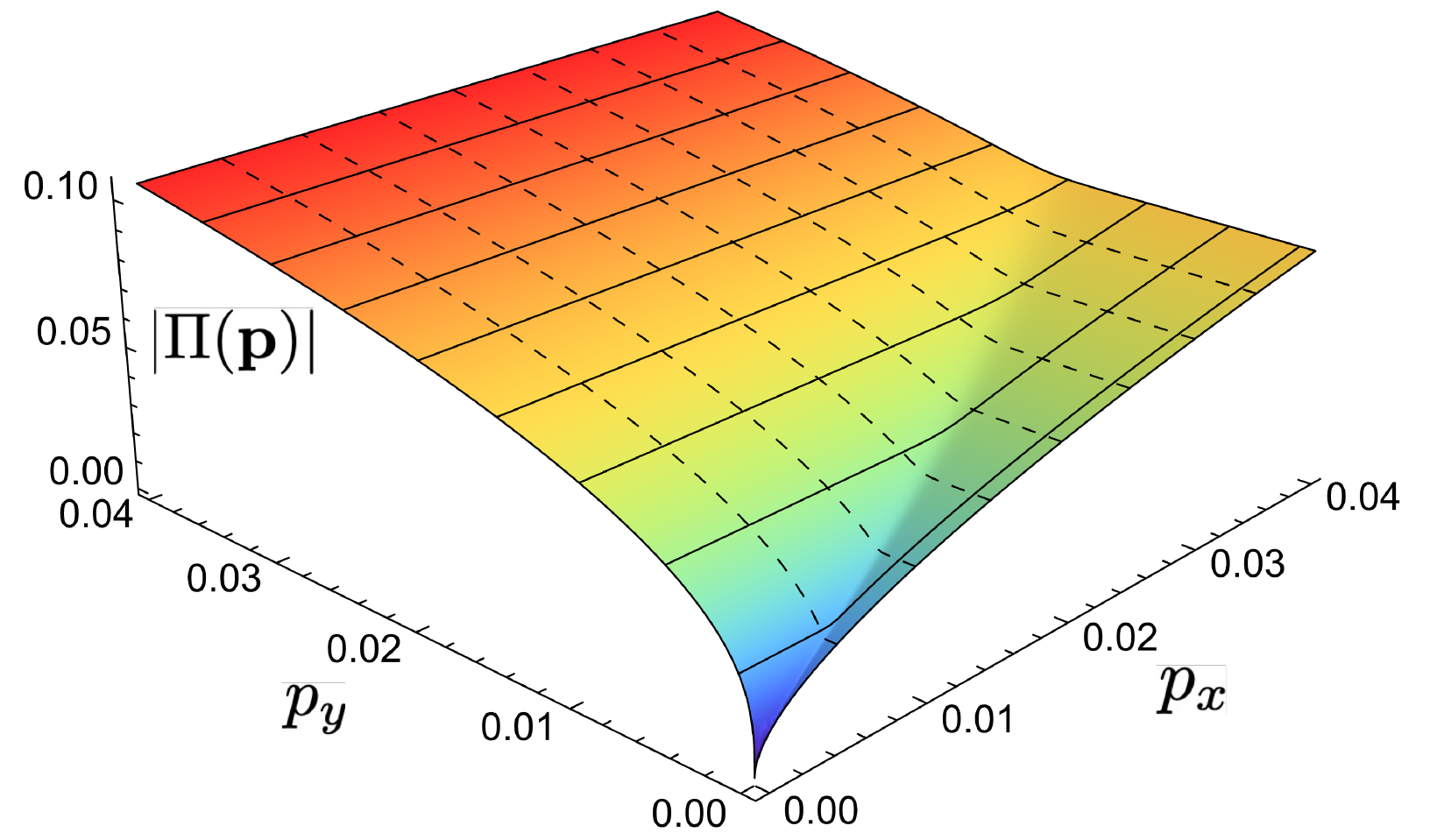}
   \caption{Static polarization function for type-II semi-Dirac fermions. The ravine residing on $p_y=(c_1/c_2)^3 p_x^{3/2}$ demarcates two regions where the polarization is dominated by $\sqrt{p_x}$ and $p_y^{1/3}$ dependence, and flat in the other direction. The dashed and solid mesh lines at constant $p_x,\,p_y$ exhibit this structure. Values of $c_1= N(0.098)= 0.392$ and $c_2=N( 0.077)=0.308$ were calculated, and $w=50$ was used for illustration.}
  \label{fig:polarization}
\end{figure}

{\it Beyond Hartree-Fock: Random Phase Approximation and static polarization function---}Going beyond the first order perturbative level, we introduce static screening via the RPA.
We have evaluated the zero frequency polarization function numerically, as outlined in the SM, and have found that it is well-approximated by the  form:
\begin{equation}
\vert\Pi(\mathbf{p})\vert = \left [  (c_1 |p_x|^{1/2})^{w} + (c_2|p_y|^{1/3})^{w} \right ]^{1/w} \! \! , \ w \gg 1
\label{eq:pol}
\end{equation}
equivalent to the following structure:
\begin{equation}
\vert\Pi(\bf{p})\vert=\begin{dcases} 
      c_1 |p_x|^{1/2}, & \frac{c_2 |p_y|^{1/3}}{c_1 |p_x|^{1/2}} \leq 1 \\
    c_2 |p_y|^{1/3}, & \frac{c_2 |p_y|^{1/3}}{c_1 |p_x|^{1/2}} \geq 1 
   \end{dcases}
\end{equation}
The number of fermionic species is taken to be $N=4$. The boundary between the two regions is visible in the plot of the polarization in Fig. \ref{fig:polarization}. We are pleasantly surprised to find that it is possible to evaluate the polarization along the linear direction $\Pi(0,p_y)$ analytically (by making use of Feynman parametrization), as described in the SM. The analytical and numerical results for the coefficient $c_2$ and $p_y^{1/3}$ dependence are in precise agreement.  

\begin{figure}
\centering
 \includegraphics[width=1\linewidth]{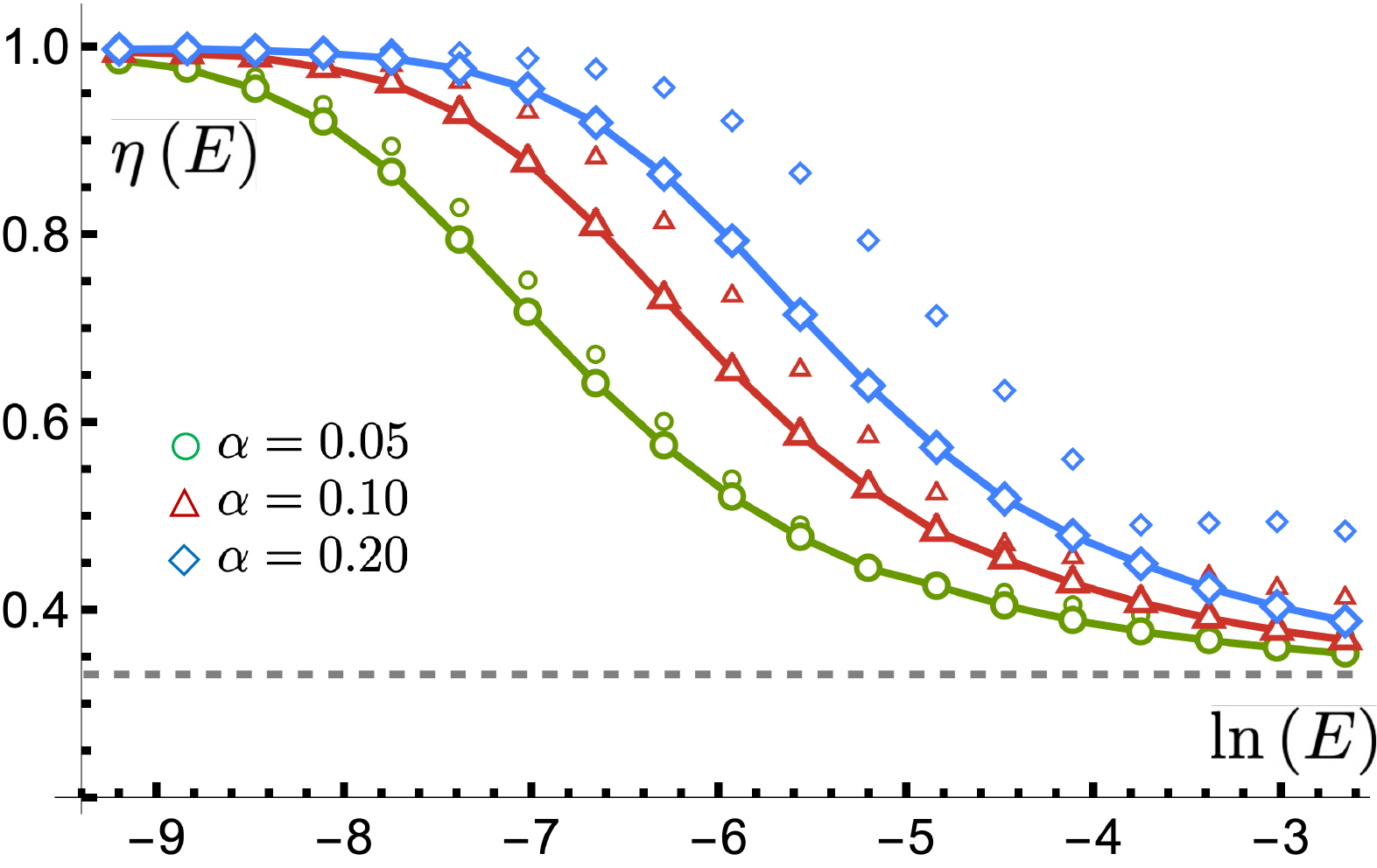}
  \caption{Evolution of the density of states scaling $\eta(E)$ under the RPA screened Coulomb interaction (solid curves) for $g=1$. We have included the corresponding Hartree-Fock results, given by the smaller unconnected markers, for comparison. For the interaction strengths $\alpha=0.05,0.10,0.20$ values of $\Delta/\varepsilon_0=1.83/(4\pi),1.55/(4\pi), 1.19/(4\pi)$ and $c/v=1.39/(4\pi),1.19/(4\pi), 0.93/(4\pi)$ are calculated respectively in the RPA. At the HF level the values $\Delta/\varepsilon_0=1.89/(4\pi)$ and $c/v=1.71/(4\pi)$ ($g=1$) are independent of $\alpha$. Results are qualitatively similar, with the smaller values of $\eta$ under the RPA indicating a weaker renormalization as expected.}
  \label{fig:dosRPA}
\end{figure}

The self-energy is then calculated as in Eq.~\eqref{se}, but instead using the dressed Coulomb potential $V_{RPA}({\bf p} ) = V({\bf p} )/(1-V({\bf p}) \Pi({\bf p}))$, and
the variation in the exponent of the renormalized density of states is summarized in Fig. \ref{fig:dosRPA}. While results are qualitatively similar to their HF counterparts, it is evident that introducing screening naturally inhibits the bare interaction effects, suppressing enhancement of the spectrum. 
The logarithmic corrections to all couplings that appear prominently within HF disappear quickly in RPA, as alpha increases, reflecting the lifting of the bare Coulomb interaction--generated infrared divergence. The RPA potential is highly anisotropic and exhibits fractional powers in both momentum directions, which dominate over the bare Coulomb potential and generally lead to generation of non-divergent terms with different symmetries,  as outlined in the SM. 
Note that $\Delta$ and $c$ are now functions of $\alpha$, and in fact decay with increasing interaction strength, partially neutralizing the growth of the crossover energy scale with $\alpha$. Details of the numerical self-energy fits are given in the SM.

{\it Conclusions and implications for observables---}The change of electronic dispersion has profound implications for the behavior of various characteristics at low temperatures/energies. 
As the density of states evolves between powers $\eta=1$ at the lowest energies and $\eta=1/3$ for higher energies (non-interacting quasiparticles), the specific heat evolves  from $C_V(T) \sim T^2$ ($\eta=1$), to 
$C_V(T) \sim T^{4/3}$ ($\eta=1/3$). In general: $C_V(T) \sim T^{\eta+1}$.

The electronic compressibility $\kappa$ is defined as: $\kappa^{-1}=n^2 (\partial \mu /\partial n)$ 
 and can be probed experimentally \cite{Girvin2019,Kotov2012,Martin2007,Sheehy2007}. Here $\mu$ is the chemical potential, $n$ the electronic density  ($n \sim \mu^{\eta +1}$), and $\partial \mu /\partial n$ is the inverse density of states.
 The charge response is typically determined by $\partial \mu /\partial n$ which is related to the inverse capacitance. 
 One finds $\partial \mu /\partial n \sim  1/\mu \sim 1/\sqrt{n}$ ($\eta=1$), and $\partial \mu /\partial n \sim  1/\mu^{1/3} \sim 1/n^{1/4}$ ($\eta=1/3$).  In general: $\partial \mu /\partial n \sim  1/\mu^{\eta} \sim 1/n^{\eta/(\eta+1)}$. 
 
 Landau level quantization \cite{Girvin2019,Abrikosov1988,Mikitik1999,Goerbig2017,deGail2012}  is governed by the semi-classical formula $A(\varepsilon_n(B))=\int_0^{\varepsilon_n(B)} \rho(\omega) d\omega \sim nB$, where $A(\varepsilon) = \int_{\varepsilon({\bf k}) \leq \varepsilon} d^2 {\bf k}/(2\pi)^2$
 is the area enclosed by the energy dispersion in momentum space. Thus we find $\varepsilon_n(B) \sim \pm  [nB]^{1/(\eta+1)}$. In particular for the non-interacting fermions, $\varepsilon_n(B) \sim \pm  [nB]^{3/4}$ presents a unique signature for experimental detection. The scaling is distinct from the conventional $B^{2/3}$ in type-I semi-Dirac systems, and there is a zero energy Landau level due to the finite Berry phase. Note that the conventional magnetic field dependence may be mimicked by interacting type-II fermions under suitable conditions.
 
Transport coefficients, such as  dynamical conductivity, can be sensitive to the geometry of the Fermi surface \cite{Maslov2017}. In our case the shape evolves from convex to concave (boomerang) with increasing Fermi energy (Fig. \ref{fig:isosurfHF}), with an expected associated change in the conductivity's frequency dependence.  Moreover, it is of direct relevance that
 nonlinear magnetotransport has been observed to be very sensitive to changes in
Fermi surface geometry in 2D semimetals such as WTe$_2$ \cite{he2019}. The change of convexity with Fermi level induces a sign inversion of the non-linear magnetoresistance, and accompanying current density response.
In the context of our study the features of the band structure are driven by interactions, and the chemical potential can easily be manipulated by doping; whereas in the magnetotransport experiments the pertinent features are electron-hole pockets, and the Fermi level is controlled via temperature. However, given the common underlying fermiology, it is reasonable to expect that peculiar magnetotransport and other potentially unusual phenomena may be observed in type-II semi-Dirac systems. 

We expect that the conclusions reached in this work will remain valid for finite, small particle density
(i.e. small filling of the upper band). Similarly to graphene the charge is not renormalized, as reflected in the non-singular behavior of the polarization function.  Topological characteristics, such as the Chern number and the anomalous Hall conductivity, are not affected by the interactions.
The interacting Dirac liquid exhibits  remarkable evolution of the Fermi surface shape as a function of the long-range interaction and Fermi energy. 
The system hosts hybrid quasiparticles with both Dirac and semi-Dirac qualities as reflected in the density of states evolution from linear to power one-third, with accompanying changes to the scaling of associated observables. 

Can these effects be detected and exploited in real materials? We would suggest, for example,  that the modification of the interaction strength $\alpha$ by using dielectric screening (i.e. $e^2 \rightarrow e^2/\kappa$, where $\kappa$ is the dielectric constant) can control the density of states exponent presented in Figs.~(\ref{fig:dosHF}), (\ref{fig:dosRPA}), thus presenting a pathway to tuning the Fermi level at which the surface changes geometry. Given the striking reversal of the current density response in WTe$_2$ for instance, we expect that such tunability might lend itself well to a variety of applications. Furthermore, our results can explain the small low-energy linear regime in (TiO$_2$)$_5$(VO$_2$)$_3$, and more generally, may be used to determine the effective interaction strength in such systems. As an illustration, we calculate an estimate of  $\alpha_{\text{eff}}$ that lies in the weak-coupling regime and is consistent with literature on the material. 

{\it Acknowledgements---}We thank Dmitrii Maslov and Bruno Uchoa for stimulating discussions. Partial financial support provided by the University of Vermont is gratefully acknowledged.

\nocite{apsrev42Control}
\bibliographystyle{apsrev4-2}
\bibliography{references}

\onecolumngrid
\newpage

\section*{Supplemental Material for “Interacting type-II semi-Dirac quasiparticles”}

\begin{center}
\textbf{ Mohamed M. Elsayed, Taras I. Lakoba, Valeri N.  Kotov}
\end{center}

\noindent
In this supplement we provide additional information and details. We have used the conventional many-body techniques,
namely the Hartree-Fock and Random Phase Approximations, as described in Ref.[S1] for example.

\section*{Hartree-Fock Approximation}

In the Hartree-Fock (HF) approximation (that is, to first order perturbatively in the interaction strength $\alpha$), the self-energy is given by Eq.~(6) of the main text, leading to 
the logarithmic renormalization of the couplings described in Eqs.~(7, 8). 
We have set the dimensionless anisotropy $g=1$ in our formulas below, in order to avoid cumbersome notation.

There are several ways to extract the leading behavior. For example for small
external momenta $k\rightarrow0$ we can expand: 
$\frac{1}{|{\bf k}-{\bf k'}|}=\frac{1}{k'}\left\{ 1+\frac{{\bf k}.{\bf k'}}{k'^{2}}-\frac{{k}^{2}}{2k'^{2}}+\frac{3({\bf k}.{\bf k'})^{2}}{2k'^{4}}\right\} +O(k^{3}).$ From here we can extract the leading infrared behavior at small momenta relative to the ultraviolet momentum cutoff $\Lambda$ ($k \ll \Lambda$): 
\begin{equation} 
\Sigma_v(k) \approx \frac{e^2}{4\pi} \int_{k'>k}  \! \! d^{2}k' \frac{{k'_y}^2}{k'^3|\varepsilon({\bf k'})|} = \frac{e^2}{4\pi} \int_{k'>k} \! \!  d^{2}k' \frac{{k'_y}^2}{k'^3|vk'_y|} = \frac{\alpha}{4\pi} \int_{k}^{\Lambda} \frac{dk'}{k'} \int_{0}^{2\pi} d\varphi \frac{\sin^2{\varphi}}{|\sin{\varphi}|}=
\frac{\alpha}{\pi}\ln{(\Lambda/k)}, 
\end{equation}
In the intermediate integration we have used the small momentum approximation $|\varepsilon({\bf k'})|\approx |vk'_y|$. Since the integral is highly anisotropic, the above expression should only be viewed as a way of intermediate momentum power counting, leading to an expected  logarithmic structure in the final result. 

 A more accurate way to extract the logarithmic behavior would be to use the energy-angle variables introduced in Eq.~(4) of the main text. This mapping fully preserves the correct anisotropic structure of the energy dispersion at low energies. We calculate the Jacobian of the transformation in the low-energy limit $|J(E,\varphi)|=\frac{E^{1/3}}{3|\sin{\varphi}|^{2/3}}$ and obtain:
 \begin{equation} 
\Sigma_v(\omega) = \frac{\alpha}{12\pi}\int_{\omega}^{\Lambda}\frac{dE}{E} I_{v}(E).
\label{sup:sigmav}
\end{equation}
Here $E,\omega, \Lambda$ are measured in units of $\varepsilon_0$ (Eq.~(4), main text).
The function $I_{v}(E)$ is defined as:
 \begin{equation} 
I_{v}(E) = 2\int_{0}^{\pi}d\vf \frac{E^{1/3}}{[\sin{\varphi}]^{2/3}}
 \frac{(E^{2/3}[\sin{\varphi}]^{2/3}-E \cos{\varphi})^2}{\left(E^{2/3}[\sin{\varphi}]^{2/3}+(E^{2/3}[\sin{\varphi}]^{2/3}-E \cos{\varphi})^2\right)^{3/2}}.
\end{equation}
We need the behavior of this function as $E \rightarrow 0$.
First, we observe that the integration over angles from $0$ to $2 \pi$ is exactly twice the integral from $0$ to $\pi$ by parity of the integrand. Next, it is clear that the value of the integral is accumulated near $\varphi \approx 0, \pi$; with the result being identical near those two values due to the symmetry of the integrand as $E\to 0$. Consequently the asymptotic behavior is determined by the following expression, for $\vf$ near the lower limit: 
\begin{eqnarray} 
&I_{v}(E)& = 4\int_{\sqrt{E}} d\vf 
 \frac{E^{5/3}\varphi^{2/3}}{\left(E^{2/3}\varphi ^{2/3}+E^{4/3}\varphi^{4/3}\right)^{3/2}}\\ \nonumber
 &&= 12/\sqrt{1+E} = 12, \ \  E \rightarrow 0.
\end{eqnarray}
Here we have used, for low energy and $\vf \sim 0$, the approximation $E^{2/3}(\sin{\vf})^{2/3}\gg E\cos{\vf}$, leading to
the condition for the angle $\vf > \sqrt{E}$.
From Eq.~(\ref{sup:sigmav}) we finally obtain, in the limit $\omega/\Lambda \ll 1$
\begin{equation} 
\Sigma_v(\omega) = \frac{\alpha}{\pi}\ln{\left(\frac {\Lambda}{\omega}\right)}, \  \ \omega = \varepsilon({\bf k}),
\label{sup:sigmavresult}
\end{equation}
where the energy dependence is to be taken ``on-shell",  i.e. at the non-interacting energy dispersion, naturally capturing the anisotropy in the system. 

Proceeding to the renormalization of the coupling $g_2$, we obtain:
\begin{equation} 
\Sigma_{g_2} \approx \frac{3e^2}{4\pi} \int_{k'>k}  \! \! d^{2}k' \frac{{k'_x}^2 {k'_y}^2}{k'^5|\varepsilon({\bf k'})|} =
\frac{\alpha}{4\pi}\int_{\omega}^{\Lambda}\frac{dE}{E} I_{g_2}(E),
\end{equation}
 where
  \begin{equation} 
I_{g_2}(E) = 2\int_{0}^{\pi}d \vf \frac{E^{1/3}}{[\sin{\varphi}]^{2/3}}
 \frac{E^{2/3}[\sin{\varphi}]^{2/3}(E^{2/3}[\sin{\varphi}]^{2/3}-E \cos{\varphi})^2}{\left(E^{2/3}[\sin{\varphi}]^{2/3}+(E^{2/3}[\sin{\varphi}]^{2/3}-E \cos{\varphi})^2\right)^{5/2}}
\end{equation}
From here, following the procedure we used in the calculation of $I_v(E)$ above, we find for $E \rightarrow 0$
\begin{eqnarray} 
&I_{g_2}(E)& = 4\int_{\sqrt{E}} d\vf 
 \frac{E^{7/3}\varphi^{4/3}}{\left(E^{2/3}\varphi ^{2/3}+E^{4/3}\varphi^{4/3}\right)^{5/2}}\\ \nonumber
 &&= 4/(1+E)^{3/2} = 4, \ E \rightarrow 0.
\end{eqnarray}
Consequently:
\begin{equation} 
\Sigma_{g_2}(\omega) = \frac{\alpha}{\pi}\ln{\left(\frac {\Lambda}{\omega}\right)}, \  \ \omega = \varepsilon({\bf k}).
\label{sup:sigmag2result}
\end{equation}

We have implemented  a full, discretized  numerical integration   of the self-energy by using the
following representation (equivalent, via change of variables, to Eq.~(6) of the main text):
$\hat{\Sigma}({\bf k})=\frac{1}{2}\int\frac{\text{d}^{2}k'}{(2\pi)^{2}}V({\bf k'})
\frac{\mathcal{H}(\mathbf{k'}+\mathbf{k})}{|\varepsilon({\bf k'}+\mathbf{k})|}.$ Here $V({\bf k})=2\pi e^2/k$. This has confirmed the 
asymptotic results, Eqs~(\ref{sup:sigmavresult}),(\ref{sup:sigmag2result}), and in addition has determined similar behavior
for the coupling $g_1$,
\begin{equation} 
\Sigma_{g_1}(\omega) = \frac{\alpha}{\pi}\ln{\left(\frac {\Lambda}{\omega}\right)}.
\label{sup:sigmag1result}
\end{equation}

Thus in the HF approximation the leading low-energy correction to the three couplings is the divergent logarithmic renormalization with identical coefficients. The rest of the self-energy contains two important non-universal (i.e. ultraviolet cutoff
 $\Lambda$ dependent) contributions $\hat{\Sigma}(0)=\alpha\Delta\,\hat{\sigma}_x$, and a leading linear term $\alpha ck_x\, \hat{\sigma}_y$ generated by the lowest order in the expansion of the potential, that we extract numerically. We find that $\Delta$ grows rapidly with $\Lambda$, whereas $c$ exhibits a much weaker increase. Both quantities show a mild dependence on $g$. The values of $\Lambda$ and $g$ used in the HF results in the main text are $\Lambda=1.1$ in units of $\varepsilon_0$ and $g=1/2$.

\section*{Polarization Function}

We calculate the static polarization function by using the conventional formula for the one-loop integral:
\begin{equation}
\Pi({\bf p})= -iN \int \int \frac{d^2k}{(2\pi)^2} \frac{d \omega}{2\pi} \ {\mbox{Tr}}(\hat{G}({\bf k},\omega)
\hat{G}({\bf k}+{\bf p},\omega)).
\end{equation}
The integer $N$ represents the number of fermionic species contributing to the vacuum polarization.
We have used $N=4$ appropriate for the material from Ref.[S2].
From here we obtain:
\begin{equation}
\Pi({\bf p})= -N\int \frac{d^2k}{(2\pi)^2} \frac{1}{\varepsilon({\bf k})+\varepsilon({\bf k}+{\bf p})}\left ( 1 - \frac{\varepsilon_{x}({\bf k})\varepsilon_{x}({\bf k}+{\bf p})+\varepsilon_{y}({\bf k})\varepsilon_{y}({\bf k}+{\bf p})}{\varepsilon({\bf k})\varepsilon({\bf k}+{\bf p})}\right ),
\label{sup:pol}
\end{equation}
with the functions $\varepsilon_{x}({\bf k})=k_x^2-k_y, \varepsilon_{y}({\bf k})=k_x k_y$, and $\varepsilon({\bf k})=\sqrt{\varepsilon_{x}({\bf k})^2+\varepsilon_{y}({\bf k})^2}$. (We have taken all the couplings to one for simplicity of notation.) We have evaluated the function Eq.~(\ref{sup:pol}) numerically which leads to the result in Eqs.~(11,12) and Figure (5)
of the main text.

It is also instructive to evaluate certain limits analytically, if possible.
It turns out that we can perform an asymptotically exact calculation of the polarization function in the $p_y$ direction.
Using the Feynman parametrization, $\frac{1}{AB}=\int_0^1\frac{dt}{\left[t(A-B)+B\right]^2}$, we obtain:
\begin{equation}
    \Pi(\mathbf{p})=\frac{-N}{8\pi^2}\int_0^1\text{d}t\int\text{d}^2\mathbf{k}\frac{t\left[\varepsilon^2(\mathbf{k}+\mathbf{p})-\varepsilon^2(\mathbf{k})\right]+\varepsilon^2(\mathbf{k})-\varepsilon_x(\mathbf{k})\varepsilon_x(\mathbf{k}+\mathbf{p})-\varepsilon_y(\mathbf{k})\varepsilon_y(\mathbf{k}+\mathbf{p})}{\left[t\left[\varepsilon^2(\mathbf{k}+\mathbf{p})-\varepsilon^2(\mathbf{k})\right]+\varepsilon^2(\mathbf{k})\right]^{3/2}}.
\end{equation}
Performing the integral over $k_y$ exactly we have
\begin{equation}
    \Pi(0,p_y) = \frac{-N}{8\pi^2}\int_0^1\text{d}t\int_{-\infty}^{\infty} \text{d}k_{x} \frac{4p_y^2t(1-t)(1+k_x^2)^{3/2}}{k_x^6+p_y^2t(1-t)(1+k_x^2)^{2}}.
\end{equation}
Let us introduce the notation $z=p_y^2t(1-t)$. We evaluate the integral for $z \ll 1$ with the result
\begin{equation}
2\int_{0}^{\infty} \text{d}k_{x} \frac{4z(1+k_x^2)^{3/2}}{k_x^6+z(1+k_x^2)^{2}}
= 8z^{1/6}\int_{0}^{\infty} \text{d}k_{x} \frac{(1+z^{1/3}k_x^2)^{3/2}}{k_x^6+(1+z^{1/3}k_x^2)^{2}} \approx \frac{8\pi}{3}z^{1/6} + O(\sqrt{z}).
\end{equation}
Along the way we have changed  variables $k_x/z^{1/6} \rightarrow k_x$, and used the integral 
$\int_{0}^{\infty} \text{d}k_{x} 1/(k_x^6+1) = \pi/3$. The final result is
\begin{equation}
    \Pi(0,p_y) = \frac{-N}{8\pi^2}\frac{8\pi}{3} p_y^{1/3} \int_0^1[t(1-t)]^{1/6}  \text{d}t=\frac{-N}{24}\frac{\Gamma(1/6)}{\Gamma(5/6)\Gamma(1/3)}p_y^{1/3}= -N \ 0.077 \ p_y^{1/3}
\end{equation}
This result is in complete agreement with the  numerical expression, Eqs.~(11,12) of the main text.

\section*{Random Phase Approximation (RPA)}
The random phase approximation takes into account the effective screening of the bare Coulomb potential. 
Within the RPA, Ref.[S1], we have for the self-energy
\begin{equation}
\hat{\Sigma}({\bf k})=\frac{1}{2}\int\frac{\text{d}^{2}k'}{(2\pi)^{2}}V_{RPA}({\bf k'})
\frac{\mathcal{H}(\mathbf{k'}+\mathbf{k})}{|\varepsilon({\bf k'}+\mathbf{k})|},
\label{sup:se}
\end{equation}
where the statically screened potential is given by 
$V_{RPA}({\bf k'}) = V({\bf k'})/(1-V({\bf k'}) \Pi({\bf k'}))$. 
The zero momentum self-energy
\begin{equation}
\hat{\Sigma}(0)=\frac{1}{2}\int\frac{\text{d}^{2}k'}{(2\pi)^{2}}V_{RPA}({\bf k'}) 
\frac{{k'}_{x}^2}{|\varepsilon({\bf k'})|}\;\hat{\sigma}_x=\alpha\Delta(\alpha)\,\hat{\sigma}_x
\label{sup:delta}
\end{equation}
now acquires a non-multiplicative dependence on $\alpha$ via $V_{RPA}$. We find that the same applies to the constant $c$ generated by the linear term $\alpha c(\alpha) k_x\,\hat{\sigma}_y$ in the small momentum expansion. 

All calculations within RPA were performed numerically
because the anisotropic form of the polarization does not allow for any analytical calculation.
The computation was performed on a disk in momentum space, $|{\bf k'}| \le \Lambda$, and was challenging because
of the presence of the small quantity $|{\bf k}|$ in \eqref{sup:se}, which led to particularly non-uniform behavior of the integrand in both the radial and azimuthal directions of ${\bf k'}.$
We then implemented a highly accurate empirical fit, which depended on both $|{\bf k}|$ and arg$({\bf k'})$,  to the numerically computed self-energy for a wide range of $\alpha$ and $\Lambda$.
We find that it takes an astonishing 36 terms to achieve a visually satisfactory fit for both $\Sigma_x$ and $\Sigma_y$ ($\hat{\Sigma}=\Sigma_x\hat{\sigma}_x+\Sigma_y\hat{\sigma}_y$).
The main effect of the accounting for polarization in 
$V_{RPA}({\bf k})$ that we have observed is a very rapid 
decrease of the leading logarithmic contributions, given by
Eqs.~\eqref{sup:sigmavresult}, \eqref{sup:sigmag2result}, and \eqref{sup:sigmag1result},
to the self-energy. For example, already for $\alpha$ as small as 0.025, these contributions have decreased more than tenfold compared to the HF approximation. 
This lifting of the infrared divergence is due to the highly anisotropic RPA potential with fractional powers in both momentum directions dominating over the bare Coulomb potential. Therefore, for the purposes of describing the crossover and evolution of the density of states, most of the aforementioned terms in the empirical fits can safely be neglected, and we may adequately represent the self-energy by retaining only few dominant terms.
For example, at the value of the Coulomb coupling
$\alpha = 0.2$ we find that the most important terms have the form:
\begin{eqnarray}
&\hat{\Sigma}_x &=\alpha\left( 0.57 k_y + \Delta(\alpha)\right)\hat{\sigma}_x \\  
&\hat{\Sigma}_y &= \alpha\left( 0.56 k_x k_y + c(\alpha) k_x\right)\hat{\sigma}_y
\label{sup:rparesults}
\end{eqnarray}
Note that the fully precise fits including all terms were used in generating the results in Fig. (6). 

\section*{Scaling of density of states}
In order to compute the energy-dependent scaling $\eta$ of the interacting density of states $\rho(E)\sim E^{\,\eta(E)}$, we first discretize the isoenergy contours into a set of points, and calculate the area enclosed by using the relation $A=(1/2)\sum_{i=1}^n(x_i y_{i+1}-x_{i+1}y_i)$, which gives the area $A$ bounded by a polygon with vertices $(x_i,y_i)$. We then perform a numerical differentiation as we increment $E$ in the interval $[10^{-4},10^{-1}]$, with equal spacing on a logarithmic scale, which returns a quantity proportional to the density of states. By taking logarithms the exponent $\eta(E)$ is easily extracted. 
 
\vspace{0.7cm}

\noindent
[S1] G. F. Giuliani and  G. Vignale, {\it Quantum Theory of the Electron Liquid} 
 (Cambridge University Press, 2010).

\noindent
[S2] H. Huang,  Z. Liu, H. Zhang, W. Duan,  and  D. Vanderbilt,  Emergence of a Chern-insulating state from a semi-Dirac dispersion, Phys. Rev. B {\bf 92}, 161115 (2015).


\end{document}